\documentclass[preprint]{elsarticle}
\usepackage{epsfig}
\usepackage{graphicx}
\usepackage{dcolumn}
\usepackage{bm}
\usepackage{graphicx}
\usepackage{dcolumn}
\usepackage{bm}
\def\br{\begin{eqnarray}}
\def\er{\end{eqnarray}}
\def\be{\begin{equation}}
\def\ee{\end{equation}}

\def\({\left(}
\def\){\right)}

\def\<{\left\langle}
\def\>{\right\rangle}

\begin{document}

\title{Light composite scalar boson from a  see-saw mechanism  in two-scale TC models.}
 
\author{A.~Doff$^1$ and A.~A.~Natale$^{2,3}$}
\ead{agomes@utfpr.edu.br$^1$,  natale@ift.unesp.br$^{2}$}
\address{$^1$ Universidade Tecnol\'ogica Federal do Paran\'a - UTFPR - DAFIS
Av Monteiro Lobato Km 04, 84016-210, Ponta Grossa, PR, Brazil \\
$^2$Centro de Ci\^encias Naturais e Humanas, Universidade Federal do ABC, 09210-170, Santo Andr\'e - SP, Brazil\\
$^3$Instituto de F\'{\i}sica Te\'orica, UNESP 
Rua Dr. Bento T. Ferraz, 271, Bloco II, 01140-070, S\~ao Paulo - SP,
Brazil  }

\date{\today}

\begin{abstract}
We consider the possibility of a light composite scalar boson arising from mass mixing between a relatively light and  heavy scalar singlets in a see-saw mechanism  expected to occur in two-scale Technicolor (TC) models. A light composite scalar boson can be generated when the TC theory features two technifermions species in different representations,  $R_1$ and $R_2$,  under a single technicolor gauge group, with characteristic scales $\Lambda_1$ and $\Lambda_2$. We determine the final composite scalar fields, $\Phi_1$ and $\Phi_2$, effective theory
using the  effective potential for composite operators approach. To generate a light composite scalar it is enough to have a walking (or quasi-conformal) behavior just for one of the technifermions representations. 
\end{abstract}



\maketitle

 
\par The nature of the electroweak symmetry breaking is one of the most important problems in
particle physics, and the $125$ GeV new resonance discovered at the LHC \cite{LHC} has many of the characteristics expected for the Standard Model (SM) Higgs boson.
If this particle is a composite or an elementary scalar boson is still an open question. Many models have considered the possibility of a light composite Higgs based on effective Higgs potentials as reviewed in Ref.\cite{bella}. The reason for the existence of the different models (or different potentials) for a composite Higgs, is a consequence of our poor knowledge of the strongly interacting theories, that is reflected in the many choices of parameters in the effective potentials. On the other hand the 
composite scalar boson mass can be calculated based on the dynamics of the theory \cite{us1}, and this approach, although more complex, is more restrictive than the analysis of potential coefficients in several specific limits. Recently  Higgs see-saw models  have been proposed to explain possible deviations from the SM predictions\cite{saw1}. In this paper we consider the possibility of a light TC scalar boson arising from mass mixing between a relatively light and  heavy composite scalar singlets from a see-saw mechanism  expected to occur in two-scale TC models\cite{Lane,ff}.

\par  We will consider the formation of a light composite scalar  boson when the TC theory features two technifermion species in different representations,  $R_1$ and $R_2$,  under a single technicolor gauge group, with characteristic scales $\Lambda_1$ and $\Lambda_2$. To determine the final effective theory for scalar composite fields \cite{cjt}, $\phi_1$ and $\phi_2$, we will review a few aspects of Ref.\cite{us5}. We start presenting the effective Lagrangian derived in the Ref.\cite{us5} in the case of only one variational effective composite field $\phi$
\br
\Omega^{(\alpha )}_{R} = \int d^4x\left[\frac{1}{2} \right. \partial_{\mu}\Phi\partial^{\mu}\Phi -\frac{\lambda_{4VR}^{(\alpha)}}{4}\Phi^4 - \left. \frac{\lambda_{6VR}^{(\alpha)}}{6}\Phi^6 - ...\right]  \, .
\label{eq33}
\er
\noindent  where the final effective Lagrangian of Eq.(\ref{eq33}) comes out when we normalize the scalar field $\phi$ according to \cite{us5}  
\be
\Phi\equiv [Z^{(\alpha )}]^{-\frac{1}{2}}\phi \, .
\label{eq32}
\ee 
This normalization appears when we consider the effect of the kinetic term in the effective action \cite{us5}. 

The index $\alpha$ in Eq.(\ref{eq32}) is
related to  most general asymptotic fermionic self-energy expression for a non-Abelian gauge theory \cite{us2,dn05}:
\be 
\Sigma^{(\alpha)} (p^2) \sim \mu \left( \frac{ \mu^2 }{p^2}\right)^{\alpha}\left[1 + b g^2 (\mu^2) \ln\left(p^2/\mu^2 \right) \right]^{-\gamma\cos (\alpha \pi)}  \,\,\, ,
\label{eq4}
\ee	
describing all possible behaviors of any generic strongly interacting theory as discussed in the sequence.
For $\alpha = 1$ we obtain the form of the effective potential associated to the asymptotic self-energy behavior predicted by the
operator product expansion (OPE) \cite{pol}
\be
\Sigma^{(1)} (p^2) \sim \frac{ \mu^3 }{p^2} ,
\label{eq5}
\ee
\noindent and for $\alpha = 0$ we obtain  the corresponding result to the  following asymptotic expression
\be
\Sigma^{(0)} (p^2) \sim \mu \left[1 + b g^2 (\mu^2) \ln\left(p^2/\mu^2 \right) \right]^{-\gamma} \, .
\label{eq6}
\ee  
The self-energy vary between these two extreme expressions as we change the number of fermions in the theory and when effective four 
fermion interactions start being important \cite{tak}.

\par The asymptotic expression shown in Eq.(\ref{eq6}) was determined in the appendix of Ref.\cite{cs} and it satisfies the Callan-Symanzik equation. It has been argued that Eq.(\ref{eq6}) may be a realistic solution in a scenario where the chiral symmetry breaking is associated to confinement and the gluons have a dynamically generated mass \cite{us12,us22,us4}. This solution also appears when using an improved renormalization group approach in QCD, associated to a finite quark condensate \cite{chan}, and it minimizes the vacuum energy as long as $n_{f}>5$ \cite{mon}. In Eqs.(\ref{eq4}), (\ref{eq5}) and (\ref{eq6}) $\mu \approx \Lambda$,  where $\Lambda$ is the characteristic mass scale of the strongly interacting theory forming the composite scalar boson,  $\mu$ is the dynamical mass and is not an observable, moreover,  from the QCD experience we may expect that they are of the same order.  $g$ is the strongly interacting running coupling constant, $b$ is the coefficient of $g^3$ term in the renormalization group $\beta$ function, $\gamma= 3c/16\pi^2 b$, and  $c$ is the quadratic Casimir operator given by  $c = \frac{1}{2}\left[C_{2}(R_{1}) +  C_{2}(R_{2}) - C_{2}(R_{3})\right]$  where $C_{2}(R_{i})$,  are the Casimir operators for fermions in the representations  $R_{1}$ or $R_{2}$ that form a composite boson in the representation $R_{3}$. We will consider only $SU(N)$ theories and the different $\alpha$ values in the interval of $0$ to $1$ will correspond to different self-energy behaviors, going from the extreme walking (or almost conformal $SU(N)$ theories \cite{SD}) to the standard OPE one \cite{us5}.

\par  The couplings ($\lambda_{nVR}^{(\alpha)}$) are given respectively by \cite{us5}
\br 
 \lambda^{(0)}_{4VR} &\equiv& \lambda_{4V}^{(0)} [Z^{(0)}]^2 = \frac{Nn_{f}}{4\pi^2}[Z^{(0)}]^2 \nonumber \\
&& \times  \left[\left(\frac{1}{\beta(4\gamma - 1)} +\frac{1}{2}\right) \right. \nonumber \\
&& \left. - \frac{4\alpha}{\beta(4\gamma - 1)}\left(\frac{1}{(4\gamma - 2)} +2\gamma\right)\right] ,
\er
\br 
&& \lambda^{(0)}_{6VR} \equiv \lambda_{6V}^{(0)}[Z^{(0)}]^3 = - \frac{Nn_{f}}{4\pi^2}\frac{[Z^{(0)}]^3}{\Lambda^2{{}_{TC}}}  \,\,\, , 
\label{eq34}
\er
\par 
and 
\br 
\lambda^{(1)}_{4VR} &\equiv& \lambda_{4V}^{(1)} [Z^{(1)}]^2 = \frac{Nn_{f}}{4\pi^2}[Z^{(1)}]^2 \nonumber \\
&& \times  \left[\frac{1}{4}\left(1  + \frac{c\alpha_{{}_{TC}}}{2\pi}\right) \right.  \nonumber \\
&&  \left. - \frac{\beta}{4\alpha}\left(\gamma + \frac{c\alpha_{{}_{TC}}}{8\pi}(4\gamma + 1)\right)\right]
\er
\br
&& \lambda^{(1)}_{6VR} \equiv \lambda_{6V}^{(1)} [Z^{(1)}]^3= - \frac{Nn_{f}}{4\pi^2}\frac{[Z^{(1)}]^3}{7\Lambda^2} \,\,\, ,
\label{eq35}
\er
\noindent in these expressions \cite{us5} 
\be
Z^{(0)}  \approx \frac{4 \pi^2\beta (2\gamma - 1)}{N{n_f}}\,\, ,\,\,Z^{(1)} \approx \frac{8\pi^2}{Nn_{f}}\left(1-\beta\gamma\right)
\label{eqZ}
\ee
\noindent where we defined  $\beta = bg^2$, $\alpha_{{}_{TC}}$ is the coupling constant of the  technicolor interaction that forms the scalar composite. 

\par  Walking technicolor theories can  have fermions belonging to different technicolor  representations and, therefore, may have two different scales  with  characteristic  chiral symmetry  breaking scales $\Lambda_1(R_1) < \Lambda_2(R_2)$. In this proposal we are assuming that  technifermions  are in  the representations $R_1$ and $R_2$ under a single technicolor gauge group as described in Ref.\cite{Lane}.     
In the model proposed by Lane and Eichten, it is assumed that the TC running coupling constant is given by 
\br 
&&\hspace{-1.5cm}\alpha_{TC}(p^2) = \alpha_2  \,\,{\rm when }\,\,p > \Lambda_2 \nonumber  \\
&& \hspace{-1.5cm} \alpha_{TC}(p^2) = \alpha_1\left[1+ \beta^1_{0}\ln\left(\frac{p^2}{\Lambda^2_1}\right)\theta(p^2 - \Lambda^2_1) \right]^{-1} \!\!{\rm when }\, \Lambda_1 < p < \Lambda_2 \nonumber 
\er 
 \noindent where $\alpha_2 = \alpha(R_2) = \frac{\pi}{3C_2(R_2) }$, $\alpha_1 = \alpha(R_1) = \frac{\pi}{3C_2(R_1) }$  and  $\beta^1_{0} =  \frac{\alpha_1}{6\pi}(11N_{TC} - 4N_1)$ and $N_1$ are  technifermions doublets in the representation $R_1$.  Note that the $N_1$  and $N_2$ doublets of technifermions   belong  to the complex TC representation $R_1$ and $R_2$,  with  dimensionality $d_1 < d_2$. For a large enough  number of $N_1$  doublets it is possible to obtain $\Lambda_1(R_1) << \Lambda_2(R_2)$ \cite{Lane} (or the decay constant $F_1 << F_2$) because 
\be 
\frac{\Lambda_2}{\Lambda_1} \approx \exp{\left(\frac{6\pi}{(11N_{TC} - 4N_1)}\left[ \alpha^{-1}(R_2) - \alpha^{-1}(R_1) \right]\right)}, 
\ee 
\noindent in this case we can assume that the asymptotic technifermions self-energy behavior in  representation $R_1$  can be described  by Eq.(4), this hypothesis can be verified with the numerical results obtained in \cite{Lane}, where in the case (a) $R_2 = A_2$
(second rank antisymmetric tensor representation), $N_1 = 6$, $N_2= 2$ for $N_{TC} = 5$, and we have $F_2/F_1 \sim 7.7$. In Ref.\cite{us5} we have shown that the decay constants for the different asymptotic behavior of the self-energies (Eq.(4) $[\alpha=1]$, Eq.(5) $[\alpha=0]$) are 
given by

\be 
n_{d_i}F^2_\alpha = \left( 1 + \frac{\alpha}{2}\right)\frac{\Lambda^2_{TC}}{Z^{(\alpha)}}
\ee
\noindent where $n_{d_i}$ corresponds to the number of  doublets of technifermions in the representation $i=1,2$.  Therefore, for a two scale TC model this relationship implies 
\be 
\frac{\sqrt{N_2}F_2}{\sqrt{N_1}F_1} \approx \sqrt{\frac{2Z^{(1)}}{3Z^{(0)}}}\frac{\Lambda_2}{\Lambda_1}.
\label{eq1212}
\ee 
\noindent  Considering Eq.(\ref{eqZ}), together with the choice of parameters presented in the previous paragraph,  the above expression 
leads to $F_2/F_1 \sim 7.3$ in agreement with the numerical value described before. Therefore, for the analysis that we shall present in this work, the asymptotic expressions,  (Eq.(4) $[\alpha=1]$, Eq.(5) $[\alpha=0]$), are a good approximation for determining the scalar spectrum of these type of two scale TC models.

\par  At low energies  we have an effective theory containing two different sets of composite scalars  $\phi_1$  and $\phi_2$,  and like the ones described in Ref.\cite{Lane},  we will assume an  ETC gauge  group  containing $N_1$ technifermions doublets in the fundamental representation  $R_1 = F$, and $N_2$ technifermions doublets, assuming  $N_2 = 1$ for $R_2$ representations (2-index antisymmetric $A_2$, 2-index symmetric $S_{2}$). The phenomenology of these type of models was already described in Ref.\cite{Lane}.

\begin{figure}[ht]
\begin{center}
\includegraphics[width=0.7\columnwidth]{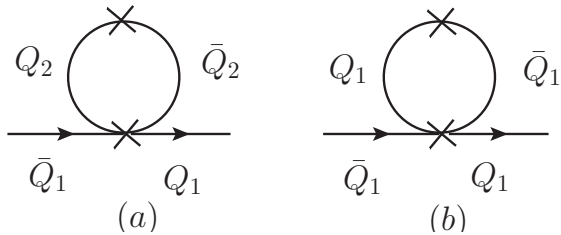}
\caption{ETC (effective four-fermion) contributions to the mixing of scalars in the representations  $R_1$ and $R_2$ }
\label{fig1}
\end{center}
\end{figure}

The fermionic content of the  model that we will discuss contain two multiplets of technifermions in the representations $R_1$ and $R_2$ of the type  
\br 
&& Q^{U}_{ETC} = \left(\begin{array}{c} {U^{a}_{R_1}}_1 \\ \vdots \\ {U^{a}_{R_1}}_i \\ {U^{a}_{R_2}}_1  \end{array}\right)_{L,R}\nonumber \,\,,\,\,Q^{D}_{ETC} =\left(\begin{array}{c} {D^{a}_{R_1}}_1 \\ \vdots \\ {D^{a}_{R_1}}_i \\ {D^{a}_{R_2}}_1   \end{array}\right)_{L,R}\\ \nonumber
\er
\noindent where $(a)$ is a technicolor index and  $(i)$ is a flavour index.  In this type of theory the ETC group would be $SU(N_ {ETC})\supset SU (N_ {TC} + N_1 + N_2)$, and  in order to incorporate the mixing between  $\phi_1 $ and $\phi_2$, we must take into account the contributions of the ETC as displayed in  Fig.(\ref{fig1}). Remembering that the self-energy can also be related to the solutions of the Bethe-Salpeter equation, we can observe that the scalar boson $\phi_1 $, formed by the fermions in the representation $R_1$ receive contributions of the condensates of the two different representations, as shown in Fig.(\ref{fig1}).

\par We can detail a little bit more the comment of the previous paragraph and the behavior of the diagrams in Fig.(\ref{fig1}). The $Q_1$ techniquarks will receive a dynamical mass due to the usual TC contribution and to the two diagrams in Fig.(\ref{fig1}), that we indicate by
\be 
\Sigma_{Q_1}(p) \approx \Sigma^{TC}_{Q_1}(p)  + \zeta\Sigma_{Q_2} + \xi\Sigma_{Q_1},
\ee 
where $\zeta$ and $\xi$ are calculable constants.
In the above expression the first one is the usual TC contribution due to condensation of $Q_1$ techniquarks in the representation $R_1$. The second comes from the ETC interaction with $Q_2$ techniquarks condensating in the representation $R_2$ and the third one is the $Q_1$ contribution  from  ETC  interactions.  Suppose now that the $Q_1$ techniquarks self-energy does not have a walking behavior, i.e. 
$\Sigma_{Q_1} (p^2)$ is given by Eq.(4), therefore the $Q_1$  ETC contribution to $\Sigma_{Q_1}(p)$,  Fig.(\ref{fig1}b) will be giving by 
\cite{dn05}
\be 
\xi \Sigma_{Q_1} \propto  O(\frac{\Lambda^3_{1}}{\Lambda^2_{ETC}}) << 1 \, ,
\ee 
which is totally negligible.

\par We can now consider the effect of $Q_2$ technifermions in the representation $R_2$. This contribution is represented by the diagram  
of Fig.(\ref{fig1}a), where we may have an extreme walking behavior for the $Q_2$ technifermions. In this case the correction due to ETC will be dominated by a self-energy of the type given by Eq.(\ref{eq6}) resulting in \cite{dn05}
\be 
 \zeta \Sigma_{Q_2} \approx \Lambda_2\left(\frac{C_{ETC}}{C_{2R_2}}\left(\frac{\alpha_{ETC}(\Lambda^2_{ETC})}{\alpha_{TC}(\Lambda^2_{ETC})} \right)^{\gamma_{2}}\right).
\label{eqdel2}
\ee
\noindent Therefore the ETC correction ($\zeta \Sigma_{Q_2}$)  plays a role similar of a bare mass term for the $\Sigma_{Q_1}(p)$ self-energy, i.e. a very hard self-energy! A similar reasoning may also be applied to the $\Sigma_{Q_2}(p) \approx \Sigma^{TC}_{Q_2}(p)  
+ \kappa \Sigma_{Q_1} $. \textit{Although only one of the technifermions representations of one given TC group has a walking behavior and
this group belongs to an ETC theory, at the end both technifermions representations will have asymptotically hard self-energies}.
In the following we will consider that the technifermions associated to the representation $R_1$ are in the fundamental representation  with a 
self-energy behaving as the one of Eq.(\ref{eq5}), and  $\Sigma_{Q_2}(p) \approx \Sigma^{TC}_{Q_2}(p)$ behaving as Eq.(\ref{eq6}) . 

\par The different terms that are going to appear in the effective action are momentum integrals of different powers of the 
self-energies $\Sigma (p)$ \cite{cjt}, which are going to be represented as $[\phi_i \Sigma_i (p)]^n$, where $\phi_i$ acts like a dynamical
effective scalar field (expanded around its zero momentum value) \cite{us5,cs}, and it is interesting to verify how it is going to be 
the behavior of the $\Sigma_i^4(p)$ term (as a function of the momentum), which is the leading term of the effective potential \cite{us5,cs}.
The fourth power of the self-energy associated to the fields $\phi_1$ and $\phi_2$, where the index $1$ will be related to technifermions with (in principle) a soft self-energy ($\alpha =1$), and the index $2$ will be related to technifermions in a representation $R_2 = S_2$ or 
$R_2 =  A_2$, with a hard self-energy ($\alpha =0$), will be written as
\br 
\Sigma^4_{1}(p^2) = && \hspace*{-0.3cm}\left( \Lambda_1 f(p)  +   a_{ETC}\Lambda_2 \right)^4 \approx \Lambda^4_1 f^4(p) + \nonumber \\
                    &&\hspace*{-0.3cm} 4a_{ETC}\Lambda^3_1\Lambda_2  f^3(p) + 6a^2_{ETC}\Lambda^2_1\Lambda^2_2  f^2(p)+...\nonumber \\
\Sigma^4_{2}(p^2) = && \hspace*{-0.3cm} \Lambda^4_2	\left[1 + \beta_{0}(R_2)\ln\left(\frac{p^2}{\Lambda_2}\right) \right]^{-4\gamma_2} \nonumber							
\er 
\noindent  where we defined $f(p) = \Lambda^2_1/p^2$ and $a_{ETC}$ is the ratio of Casimir operators and couplings of Eq.(\ref{eqdel2}).

After some lengthy calculation, that follows the same steps delineated in Ref.\cite{us5}, we obtain the following effective Lagrangian using the self-energies described previously  
\br 
\Omega(\Phi_1, \Phi_2) = \hspace*{-0.3cm}&&\int d^4x\left[\frac{1}{2}\partial_{\mu}\Phi_1\partial^{\mu}\Phi_1 + \frac{1}{2}\partial_{\mu}\Phi_2\partial^{\mu}\Phi_2  \right. \nonumber \\
&& \left. -\frac{\lambda^{R_1}_{4n}}{4}Tr\Phi_{1}^4 - \frac{\lambda^{R_2}_{4n}}{4}Tr\Phi_{2}^4 - \frac{\lambda^{R_1,R_2}_{4n}}{4}Tr\Phi_{1}^2\Phi_{2}^2 \right.\nonumber \\ 
&& \left.  - \frac{\lambda^{R_1}_{6n}}{4}Tr\Phi_{1}^6  - \frac{\lambda^{R_2}_{6n}}{4}Tr\Phi_{2}^6 \right].
\label{omfin}
\er 

\noindent In Eq.(\ref{omfin}) we included the contribution of the kinetic terms in the effective action \cite{cs}. The inclusion of these
terms lead to the normalization condition  
\be
\Phi_i = \frac{\phi_i}{Z^{1/2}(R_i)},
\ee 
and the coefficients $\lambda^{R_i}_{4,6}$  and $\lambda^{R_1,R_2}_{4}$ are the following
\br 
&& \lambda^{R_1}_{4} = \frac{N_{TC}N_1}{2\pi^2}\frac{1}{4} \nonumber  \\
&&\lambda^{R_2}_{4} = \frac{N_{TC}N_2}{2\pi^2} \left(\frac{1}{\beta(4\gamma_2 - 1 )} + \frac{1}{2}\right) \nonumber \\
&& \lambda^{R_1,R_2}_{4} = \frac{3N_{TC}N_1}{4\pi^2}\left(\frac{C_{ETC}}{C_{2R_2}}\left(\frac{\alpha_{ETC}(\Lambda^2_{ETC})}{\alpha_{TC}(\Lambda^2_{ETC})} \right)^{\gamma_{2}}\right)^2\nonumber\\
&& \lambda^{R_1}_{6} = -\frac{N_{TC}N_1}{2\pi^2} \frac{1}{7\Lambda^2_1} \nonumber \\
&& \lambda^{R_2}_{6} = -\frac{N_{TC}N_2}{2\pi^2} \frac{1}{\Lambda^2_2} \nonumber \\ 
\er
\noindent where for the representations $i=1,2$ we have
\br 
&& b_i = \frac{1}{48\pi^2}\left(11N_{TC} - 8T(R_i)N_i \right) \nonumber  \\
&& \gamma_i = \frac{3C(R_i)}{16\pi^2b_i}\nonumber \\
&& b_{ETC} =  \frac{1}{48\pi^2}\left(11N_{ETC} - 8T(R_1)N_1 - 8T(R_2)N_2 \right) \nonumber \\
&& \alpha_{ETC}(\Lambda_{ETC}) = \frac{\alpha_{ETC}(\Lambda_2)}{\left[1 + 4\pi b_{ETC}\alpha_{ETC}(\Lambda_{2})\ln\left(\frac{\Lambda^2_{ETC}}{\Lambda^2}\right) \right]} \nonumber \\
&& \alpha_{TC}(\Lambda_{ETC})\approx \alpha_{TC}(\Lambda_2) \approx \frac{\pi}{3C_2(R_2)},  
\er
\noindent in the previous expressions we  assume the MAC hypothesis and the normalized constants  $\lambda^{R_i}_{4n,6n}$ and $\lambda^{R_1,R_2}_{4n}$ are identified as
\br 
&& \lambda^{R_i}_{4n} = \lambda^{R_i}_{4}Z^{2}(R_i) \nonumber \\ 
&& \lambda^{R_1,R_2}_{4n} = \lambda^{R_1,R_2}_{4} Z(R_1)Z(R_2) \nonumber \\
&& \lambda^{R_i}_{6n} = \lambda^{R_i}_{6}Z^{3}(R_i) \nonumber \\ 
\er 
\noindent and the normalization coefficients $Z(R_i)$  are
\br 
&& Z(R_1) = \frac{16\pi^2}{N_{TC}N_1}\left(1-\beta_1\gamma_1\right)\nonumber \\ 
&& Z(R_2) = \frac{8\pi^2\beta_2(2\gamma_2 - 1)}{N_{TC}N_2}.
\er

\par The most important characteristic of this effective Lagrangian is the mixing term
\be
\lambda^{R_1,R_2}_{4} = \frac{3N_{TC}N_1}{2\pi^2}\left(\frac{C_{ETC}}{C_{2R_2}}\left(\frac{\alpha_{ETC}(\Lambda^2_{ETC})}{\alpha_{TC}(\Lambda^2_{ETC})} \right)^{\gamma_{2}}\right)^2 .
\label{mix}
\ee
This mixing is the one that defines the splitting between the effective fields $\phi_1$ and $\phi_2$, as discussed by Foadi and Frandsen 
\cite{ff}, whereas within the approach taken in this work their parameter $\delta$ \cite{ff}, characterizing the mixing  in the mass  matrix,  is
\be 
\delta = \frac{\lambda^{R_1,R_2}_{4n}}{\sqrt{\lambda^{R_1}_{4n}\lambda^{R_2}_{4n}}}.
\label{xx} 
\ee 
We emphasize that this mixing appears naturally in a two-scale TC model, where it is enough that one of the scales, and the fermionic
representation associated to it, has an extreme walking behavior and the TC group is embedded into an ETC theory. In this work we will be considering two different situations for technifermions in $R_2$ representation, case (a) [with $R_2= A_2$, $N_2=1$ , $N_1 = 10$] and 
case (b) [with $R_2= S_2$, $N_2=1$ , $N_1 = 8$]. This choice of fermionic content guarantees the preservation of asymptotic freedom and
walking behavior. 

\par  In the case of a large mixing we certainly can obtain a light scalar composite boson with a few hundred GeV mass. We show, as an example, in 
Fig.(2) the behaviour of the parameter $\delta$ in the case (a).   
\par 
\begin{figure}[ht]
\begin{center}
\includegraphics[width=0.8\columnwidth]{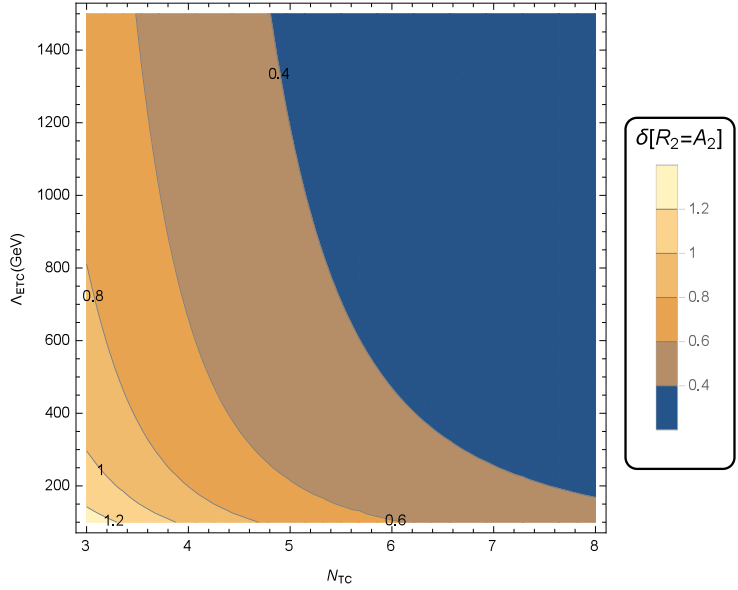}\\
\label{delta}
\caption{In this figure we show the behavior  of the mixing term $\delta$ as a function of $N_{TC}$(x-axis) and $\Lambda_{ETC}$ 
(y-axis). The figure  corresponds to the case (a),  where  $R_2 = A_2$, $N_2=1$ , $N_1 = 10$. From this figure it is possible to verify 
 that for the region compatible with the experimental limit on to Higgs mass (see Fig.(\ref{fig3})),  $\delta \approx 0.4$ and $\Lambda_{ETC} > 500 TeV$. }
\end{center} 
\end{figure}

\par Considering Eq.(11), and $F_{2}\sim 250GeV$,  we note that the scale $\Lambda_2 $ is defined by
\be 
N_2 F^2_{2}  = \frac{\Lambda^2_2}{Z^{(0)}}
\ee 
\noindent  which leads to
\be 
\Lambda_2 = \frac{2\pi F_{2}\sqrt{\beta(2\gamma - 1)}}{\sqrt{N_{TC}}} \sim  \frac{O(TeV)}{\sqrt{N_{TC}}}. 
\ee 
\noindent Finally, assuming  
\be 
M^{2}_{\Phi_i} = \frac{\partial^2\Omega(\Phi_i)}{\partial \Phi^2_i}\left|_{\Phi = \Phi_{min}}\right.
\ee 
\noindent  we obtain
\br 
&& M^{2}_{{}_{\Phi_i}} \approx 2\lambda^{R_i}_{4n}\left(\frac{\lambda^{R_i}_{4n}}{\lambda^{R_i}_{6n}}\right). 
\er

\par We can write the following mass matrix in the base formed by the composite scalars ($\Phi_1$)  and ($\Phi_2$)  
\br
M^2_{\Phi_1 ,\Phi_2} = \left(\begin{array}{cc} M^2_1 & M^2_{12}\delta \\ \delta M^2_{12} & M^2_{2} \end{array}\right).
\label{Meig}
\er 
\noindent The eigenvalues of this matrix provide the mass spectrum for the light scalar $(H_1)$ and heavy $(H_2)$, including the 
mixing effect parametrized by $\delta$, where 
\br 
&& M^2_i = 2\lambda^{R_i}_{4n}\left(\frac{\lambda^{R_i}_{4n}}{\lambda^{R_i}_{6n}}\right) \nonumber \\
&& M^2_{12} = M_1M_2 . \nonumber \\
\er 
\par
\begin{figure}[h]
\begin{center}
\includegraphics[width=0.9\columnwidth]{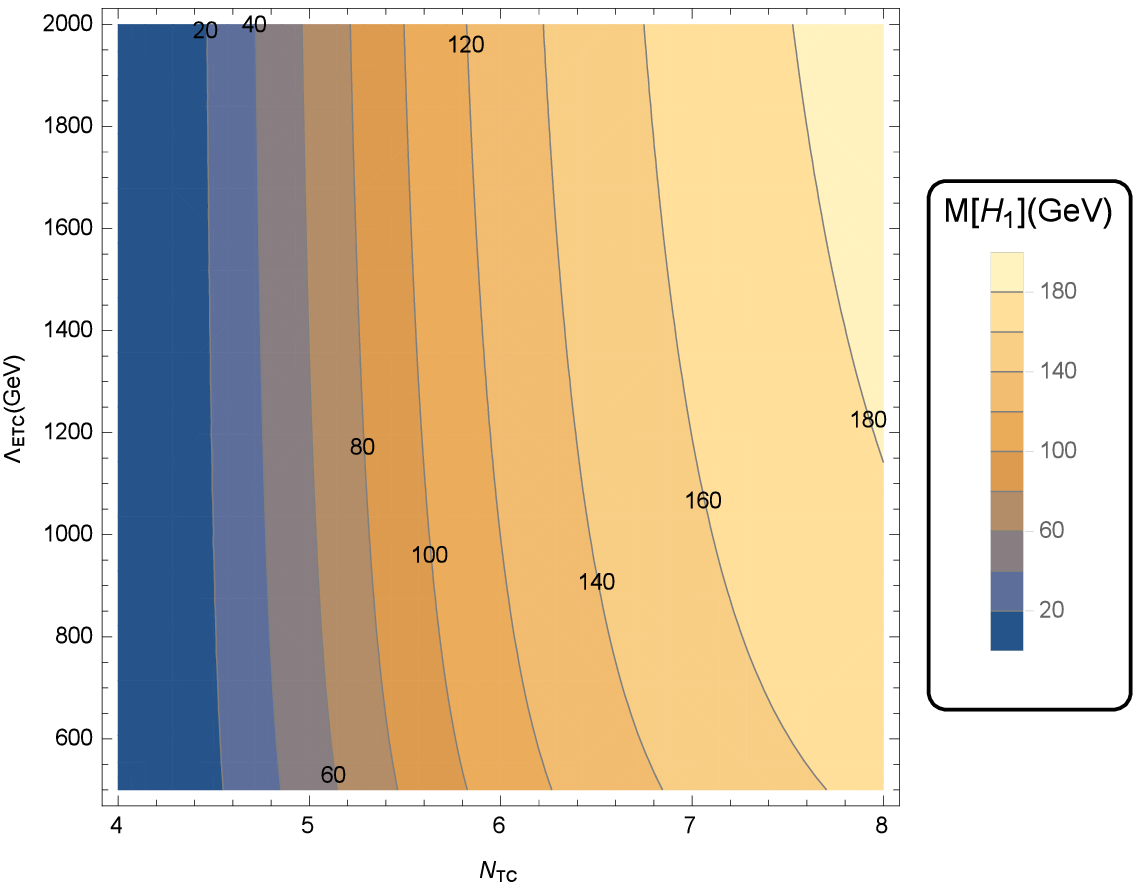}
\\
\includegraphics[width=0.9\columnwidth]{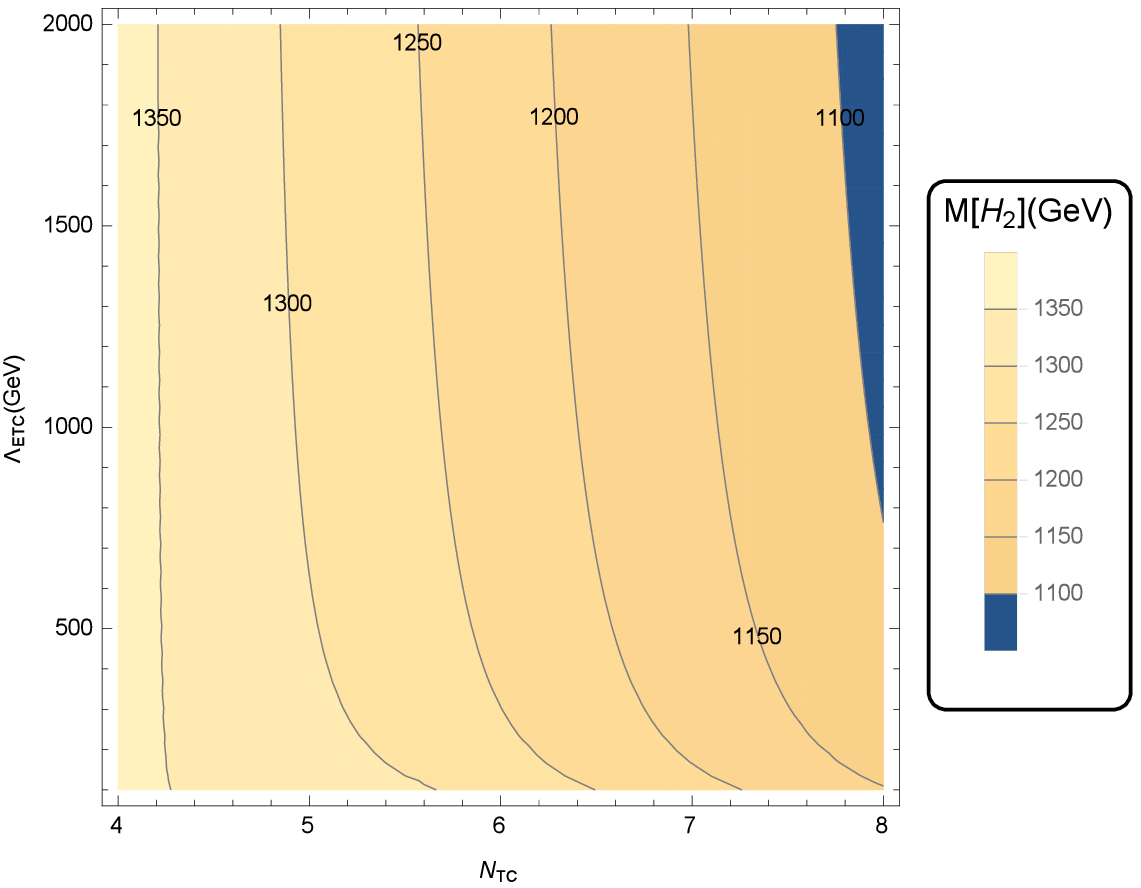} \\
\caption{The light composite scalar $H_1$ and heavier composite scalar $H_2$ regions of masses as a function of the parameters $N_{TC}$ and $\Lambda_{ETC}$ in the  
case (a), which is similar to the one considered by Lane and Eichten in Ref.\cite{Lane}.}
\label{fig3}
\end{center} 
\end{figure}

\begin{figure}[h]
\begin{center}
\includegraphics[width=0.9\columnwidth]{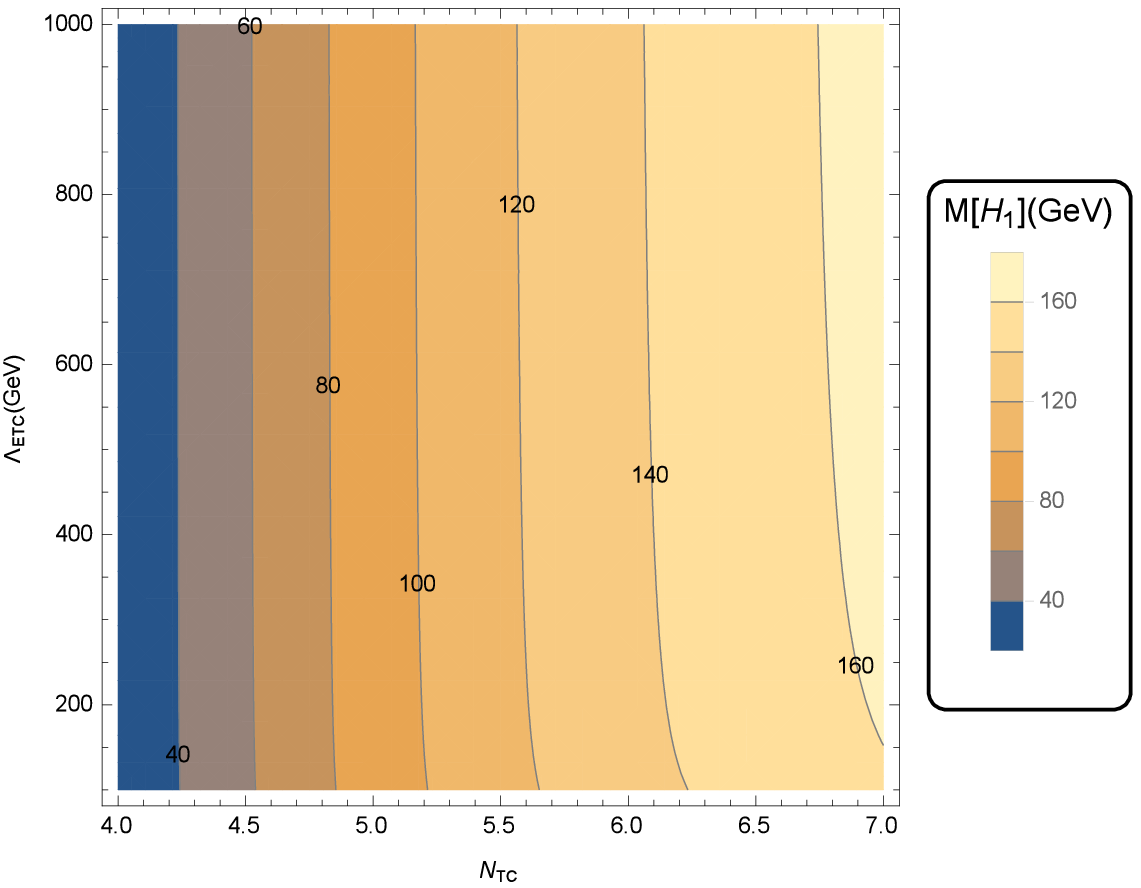}
\\
\includegraphics[width=0.9\columnwidth]{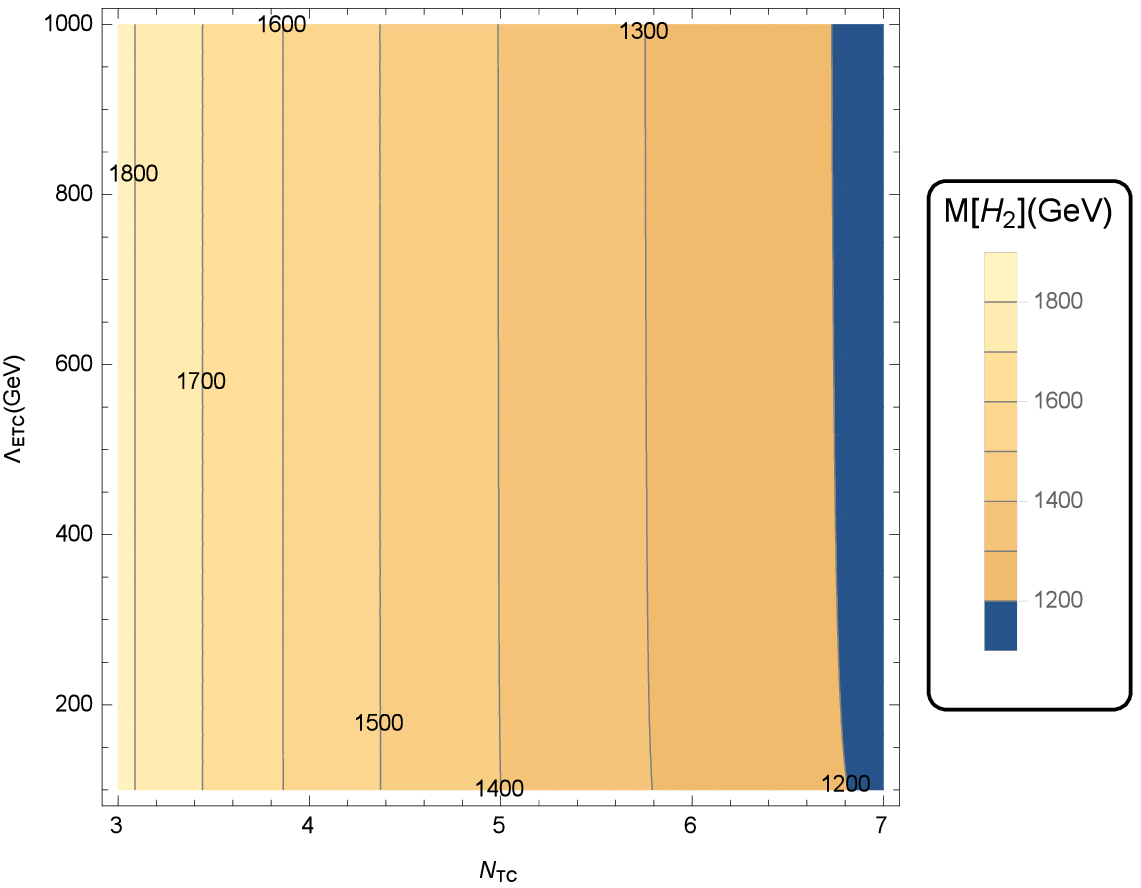} \\
\caption{Light ($H_1$) and heavy ($H_2$) scalar composite region of masses in the case (b) [$R_2= S_2$, $N_2=1$ , $N_1 = 8$] as a function of the parameters $N_{TC}$ and $\Lambda_{ETC}$. }
\label{fig4}
\end{center} 
\end{figure}

  From the above equations we can determine the mass spectrum for the scalar bosons, $H_1(R_1)$ and $H_2(R_2)$, which are the diagonalized masses of the scalars $\Phi_1$ and $\Phi_2$  and these results are shown in Figs. (\ref{fig3}) and (\ref{fig4}), where we present  the mass spectrum obtained for the  light  and  heavier  composite scalars $H_1(R_1)$ and $H_2(R_2)$ in the cases where $R_1=F$,  $R_2 = A_2$ or $R_2 = S_2$.

\par In this work  we have computed an effective action for a composite scalar boson system formed by two technifermion species in different representations,  $R_1$ and $R_2$,  under a single technicolor gauge group with characteristic scales $\Lambda_1$ and $\Lambda_2$ as the original proposal presented  in Ref.\cite{Lane}. The calculation is based on an effective action for composite operators\cite{us5}, 
the novelty of the calculus presented in this work is that  we included technifermions  in different representations,  $R_1$ and $R_2$,  under a single technicolor gauge group. Our main results are described in Figs. (\ref{fig3}) , (\ref{fig4}). The mixing between the composite scalar bosons  $\Phi_1$ and $\Phi_2$ is  responsible for generating a light scalar composite
of a few hundred GeV mass. A particular example
of the values of this mixing is shown in Fig.(2). To obtain a large mixing it is enough that one of the technifermions representations has a walking behavior and the TC group is embedded in an ETC theory. At the end the technifermions of both representations will have asymptotically hard self-energies.
 
For a set of parameters similar to the ones used in  Ref.\cite{Lane} in the case $R_2=A_2$, we obtain the same TC group necessary to generate the walking behavior, $SU(6)_{TC}$, leading to $M_{H_1} \sim {\cal{O}}(100)$ GeV . This result reinforces  the validity of  hypothesis  
discussed below Eq.(\ref{eq1212}), and this is a  consequence  of the walking (or quasi-conformal) technicolor theory. Furthermore, the large anomalous dimensions $\gamma_m$ enhance light-scale technipion masses, $M_{\pi1} > M_{\rho1} - M_{W} $, where technirho mass  $M_{\rho1} \sim 250GeV$.  The difference between the results obtained for the representations  $R_2=A_2$  and $R_2=S_2$ is that in the $A_2$ case we obtain a light  scalar mass only with a large ETC scale  $(\Lambda_{ETC} > 10^3 TeV)$.  For the heavy scalar bosons obtained with $R_2=S_2$ or $R_2= A_2$  we expect the mass to be in the range $[1200-1300]$ GeV. 

It is interesting to shortly digress the case where this light scalar composite could be
related to the $125$ GeV scalar resonance found at CERN. The observed boson has couplings to the top and bottom quarks of the order expected for a fundamental SM Higgs boson. The
fermionic couplings in a realistic composite scalar model will involve the ETC group and a delicate alignment of the $H_1$ and $H_2$ vacua, where only $H_2$ may resemble a
fundamental scalar. Our model is far away from a realistic model since we have not defined a specific ETC theory. However we can imagine a theory where the fermionic masses are
not generated as usual, by different ETC mass scales, but a horizontal symmetry is introduced, as in \cite{ux}, where the top quark (or the third fermionic generation) obtain its
mass associated to a large ETC scale, or coupling mostly to the $H_2$ scalar composite, without generating undesirable four-fermion interactions incompatible with the
experimental data. We have also to remember that when QCD is embedded into a large ETC group together with the different
TC fermionic representations, we actually will be dealing with tree different set of scales, all of them with possible hard asymptotic contributions to the self-energies due to 
the mechanism discussed here, where the horizontal symmetry will act in order to provide the desirable fermionic couplings with the different scales. Of course, a detailed model in this direction is
not easy to obtain and is out of the scope of this work.

In Ref.\cite{us4} we considered the possibility of generating  a light TC scalar boson based on the use of the Bethe-Salpeter equation and its normalization condition, as a function of the $SU(N)$ group and the respective fermionic representation. In that work
we discussed how difficult was to generate a light scalar composite; what was possible, for example in the case of fermions in the
fundamental representation, only for a specific (and large) number of fermions and moderate $N_{TC}$. In this work we discuss a different possibility for generating a light composite in a type of see saw mechanism in a two-scale TC model, and a small scalar mass is again
generated in similar conditions. It is possible that the mixing mechanism that we propose here may be extended to models with
more than one TC group, although it is also possible to envisage that in this case we shall need a more complex ETC interaction in
order to mix the different groups.

A point to be noted is that the possibility of obtaining a light composite scalar according to the approach discussed in Ref.\cite{us4}, first obtained in \cite{us7}, is that this result is a direct consequence  of extreme walking (or quasi-conformal) technicolor theories, where the  asymptotic self-energy  behavior  is described by Eq.(5), this same behavior must also be present to generate a large mixing  ($\delta$), necessary to obtain a light scalar boson mass of approximately a few hundred GeV in a two-scale model.  In this work we identified that, regardless of the approach used for generating a light composite scalar boson, the behavior exhibited by extreme walking (or quasi-conformal) technicolor theories is the main feature needed in any model to produce a light composite scalar boson.

\section*{Acknowledgments}
This research was partially supported  by the Conselho Nacional de Desenvolvimento Cient\'{\i}fico e Tecnol\'ogico
(CNPq) and by grant 2013/22079-8 of Funda\c c\~ao de Amparo \`a Pesquisa do Estado de S\~ao Paulo (FAPESP).  

\begin{center}
\noindent\rule{5cm}{0.4pt}
\end{center}


\begin{thebibliography}{10}

\bibitem{LHC} ATLAS Collaboration, Phys. Lett. B 716, 1 (2012); CMS Collaboration, arXiv:1207.7235
[hep-ex].

\bibitem{bella} B. Bellazzini, C. Cs\'aki and J. Serra, Eur. Phys. J. C \textbf{74}, 2766 (2014).

\bibitem{us1} A. Doff and A. A. Natale, Phys. Lett. B {\bf 677}, 301 (2009).

\bibitem{saw1} P. S. Bhupal Dev, Dilip Kumar Ghosh,  Nobuchika Okadac and Ipsita Sahab, JHEP \textbf{0150}, 03 (2013); 
Rabindra N. Mohapatra and Yongchao Zhanga,  JHEP \textbf{0072}, 06 (2014). 

\bibitem{Lane} K. Lane and E. Eichten, Phys. Lett. B \textbf{222}, 274 (1989).

\bibitem{ff} Roshan Foadi and Mads T. Frandsen,  arXiv:1212.0015v1.

\bibitem{cjt} J. M. Cornwall, R. Jackiw and E. T. Tomboulis, Phys. Rev. D {\bf 10}, 2428 (1974). 

\bibitem{us5} A. Doff, A. A. Natale and P. S. Rodrigues da Silva, Phys. Rev. D \textbf{77}, 075012(2008).

\bibitem{us2} A. Doff and A. A. Natale, Phys. Lett. B {\bf 537}, 275 (2002).

\bibitem{dn05} A. Doff and A.A. Natale, Phys.Rev. {\bf D68} 077702, (2003).

\bibitem{pol} H. D. Politzer, Nucl. Phys. B {\bf 117}, 397 (1976).

\bibitem{tak} T. Takeuchi, Phys. Rev. D {\bf 40}, 2697 (1989); K.-I.Kondo, S. Shuto and K. Yamawaki, Mod. Phys.
Lett. A {\bf 6}, 3385 (1991).

\bibitem{cs} J. M. Cornwall and R. C. Shellard, Phys. Rev. D {\bf 18}, 1216 (1978).

\bibitem{us12} A. Doff, F. A. Machado and A. A. Natale, Annals of Physics
{\bf  327} 1030 (2012).

\bibitem{us22} A. Doff, F. A. Machado and A. A. Natale, New. J. Phys.
{\bf 14}, 103043 (2012).

\bibitem{us4} A. Doff, E. G. S. Luna and A. A. Natale, Phys. Rev. D {\bf 88}, 055008 (2013).

\bibitem{chan} L.-N. Chang and N.-P. Chang, Phys. Rev. D {\bf 29}, 312
(1984); Phys. Rev. Lett. {\bf 54}, 2407 (1985); N.-P. Chang and D. S. Li,
Phys. Rev. D {\bf 30}, 790 (1984).

\bibitem{mon} J. C. Montero, A. A. Natale, V. Pleitez and S. F. Novaes, Phys. Lett. B {\bf 161}, 151 (1985).

\bibitem{SD} D. D. Dietrich and F. Sannino, Phys.Rev.D {\bf75}, 085018 (2007). 

\bibitem{ux} A. Doff and A. A. Natale, Eur. Phys. J. C \textbf{32}, 417 (2003).

\bibitem{us7} A. Doff, A. A. Natale and P. S. Rodrigues da Silva, Phys. Rev. D \textbf{80},  055005 (2009).

\end{thebibliography}
\end{document}